\documentclass{aastex631}
\usepackage{amsmath}
\graphicspath{{figures/}}
\newcommand{\NH}{N_{\mathrm{H}_2}}
\newcommand{\neff}{n_{\rm eff}}
\newcommand{\leff}{l_{\rm eff}}
\newcommand{\Rt}{\mathcal{R}_{\rm tide}}
\newcommand{\tff}{t_{\rm ff}}
\newcommand{\texten}{t_{\rm ext}}
\newcommand{\Lt}{L_{\rm tide}}
\newcommand{\Ah}{A_{\rm h}}
\newcommand{\Aproj}{\langle A\rangle}
\newcommand{\Ms}{M_\odot}
\begin{document}
\title{Centralized or Multipolar Collapse: Gravitational Architectures of DR21 and W75N in Cygnus-X North}
\author[0000-0003-0596-6608]{Mengke Zhao}
\affiliation{School of Astronomy and Space Science, Nanjing University, 163 Xianlin Avenue, Nanjing 210023, Jiangsu, People's Republic of China}
\affiliation{Key Laboratory of Modern Astronomy and Astrophysics (Nanjing University), Ministry of Education, Nanjing 210023, Jiangsu, People's Republic of China}
\author[0000-0003-3144-1952]{Guang-Xing Li}
\affiliation{South-Western Institute for Astronomy Research, Yunnan University, Kunming 650091, People's Republic of China}
\author[0000-0002-5093-5088]{Keping Qiu}
\affiliation{School of Astronomy and Space Science, Nanjing University, 163 Xianlin Avenue, Nanjing 210023, Jiangsu, People's Republic of China}
\affiliation{Key Laboratory of Modern Astronomy and Astrophysics (Nanjing University), Ministry of Education, Nanjing 210023, Jiangsu, People's Republic of China}
\correspondingauthor{Guang-Xing Li, Keping Qiu}
\email{ligx.ngc7293@gmail.com, kpqiu@nju.edu.cn}

\begin{abstract}
Do all massive star-forming regions collapse the same way?
Gravity is a long-range force, and the mass distribution throughout
a cloud determines its tidal field, which describes gravitational
compression and stretching.
Using the Volume Density Mapper, we reconstruct the three-dimensional
density distribution of DR21 and W75N, two neighboring regions in
Cygnus-X North, from the column-density map and calculate their tidal fields.
Tidal anisotropy measures the departure from isotropic compression
relative to the gravitational scale set by the local density.
In DR21, the main ridge organizes the tidal field, with strong
projected anisotropy extending to $\sim2.8$ pc from the ridge.
In W75N, several mass concentrations organize the field, with
corresponding extents of $\sim0.9$--$1.2$ pc.
As the column-density threshold increases, the characteristic volume
density and density exponent remain nearly unchanged in DR21 but
shift toward higher densities and more negative exponents in W75N.
Analytical filament models account for the approximate strength and
radial decline of the ridge's tidal field, whose extent depends on
the filament mass distribution and ambient density.
Despite similar total gas masses and tidal-to-free-fall timescale
distributions, DR21 has a larger fraction of its extensive column
mass associated with strong projected tidal anisotropy.
DR21 also contains more high-column-density gas and more mass
in massive cores.
These differences suggest that the spatial organization of gravity
may influence mass assembly during cloud collapse.
The tidal calculations use the publicly available GravTide3D package
(\url{https://github.com/meng-ke/GravTide3D}).

\par\smallskip
\noindent\makebox[\linewidth][c]{%
    \includegraphics[width=0.85\linewidth]{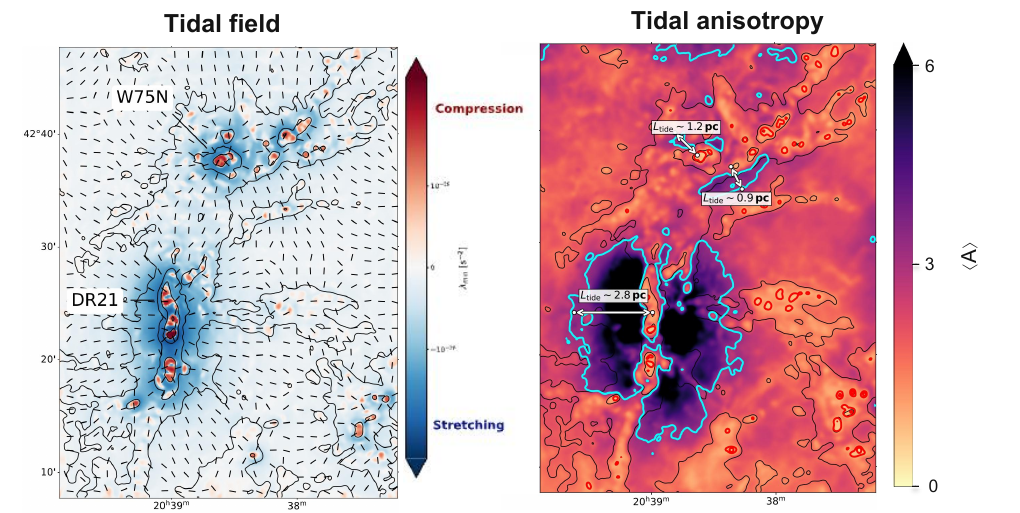}
}\par
{\small
\noindent
 \it Tidal fields characterize the differential effects of gravity, and tidal anisotropy reveals their directional dependence.
\par}
\end{abstract}
% \keywords{ISM: clouds --- ISM: structure --- gravitation --- stars: formation --- stars: massive --- methods: numerical}

\section{Introduction}\label{sec:intro}
Star formation is a complex physical process governed by gravity, turbulence, and magnetic fields \citep{Shu1987,McKee2007,Crutcher2012}. 
A central problem in massive star formation is how gas from different parts of a molecular cloud becomes concentrated into massive cores while the cloud undergoes fragmentation \citep{Zinnecker2007,Motte2018,Krumholz2019}. 
Molecular clouds exhibit hierarchical density structures, and gas can flow through filamentary networks toward dense regions \citep{Andre2014,Hennemann2012,Hacar2023}. 
In a collapsing cloud, local contraction and fragmentation take place concurrently with mass transport on larger scales \citep{VazquezSemadeni2019,Li2024c,LiZhaoLu2025}. 
The formation of massive cores thus depends on how gravity acts across different spatial scales.

Gravity is a long-range force. The gravitational field at a given location depends on the distribution of mass throughout the cloud rather than on the local density alone \citep{BallesterosParedes2009,Li2024a}. Gas at the same density can therefore be subject to different gravitational forces and different conditions for collapse. The gravitational acceleration $\boldsymbol{g}=-\nabla\Phi$ describes the acceleration of a gas element, whereas the tidal tensor $T_{ij}=\partial^2\Phi/\partial x_i\partial x_j$ describes the relative acceleration of neighboring elements. This differential acceleration causes gravitational stretching or compression. A mass concentration can attract the surrounding gas while opposing its independent contraction along a stretching direction. External tides can thus modify the local Jeans criterion \citep{Jog2013,ZavalaMolina2023,Li2024a,Li2024b} and have been linked to gas inflows and linewidths in star-forming regions \citep{Zhou2024,Zhou2025}.

The three eigenvalues of the tidal tensor describe gravitational compression and stretching along its principal axes. With our sign convention, a positive eigenvalue corresponds to compression and a negative one to stretching. A positive minimum eigenvalue $\lambda_{\min}$ indicates compression along all three axes, whereas a negative value indicates at least one stretching direction \citep{Li2024a}. Through Poisson's equation the local density fixes the sum of the three eigenvalues, but not their individual values. The differences between the eigenvalues describe the anisotropy of gravitational deformation. Tidal anisotropy is prevalent in simulations of molecular clouds \citep{Ganguly2024}, and the evolution of a perturbation depends on the tides along all three principal axes \citep{Li2024b,DumondChabrier2025}. In the low-density gas surrounding a mass concentration, anisotropic tides can be strong relative to the gravitational scale set by the local density. A mass concentration can thus influence the conditions for fragmentation well beyond its dense gas.

Cygnus-X, at a distance of $\sim1.4$ kpc \citep{Rygl2012}, is one of the nearest complexes forming massive stars and stellar clusters. Observations of Cygnus-X, including those from the CENSUS survey, have revealed its gas structure, kinematics, and core population from cloud to core scales \citep{Motte2007,Cao2019,Cao2021,LiCaoQiu2021,Wang2022,Zhang2024,LiQiu2026,XingQiu2026}. Here we reconstruct the three-dimensional tidal fields of DR21 and W75N in Cygnus-X North and compare their spatial organization. We use analytical filament models to examine the extent of the tidal field around a massive filament, and compare the density structure and core population of the two regions to explore how the organization of gravity is related to mass concentration.

\section{Data and Tidal Reconstruction}\label{sec:data}
DR21 and W75N are two neighboring massive star-forming regions
in Cygnus-X North \citep{Motte2007,Kumar2007,Rygl2012}. We use the Cygnus-X $\NH$ column-density map derived from dust-continuum SED fitting by \citet{Cao2019}. Both regions are extracted from the same parent map, regridded to $6''$ pixels, and analyzed at a common distance of 1.4 kpc. The analysis field includes the surrounding mass distribution required for the tidal calculation. Pixels have equal area, so $\NH$ provides the mass weighting used below.

\subsection{Line-of-sight density reconstruction}\label{sec:vdm}
The gravitational potential is determined by the volume density through Poisson's equation, $\nabla^2\Phi=4\pi G\rho$. The projected gravitational acceleration can be estimated from a column-density map by assuming a uniform cloud thickness \citep{He2023}. The tidal tensor, however, describes spatial variations of the acceleration in three dimensions and is more sensitive to small-scale density structure. We therefore reconstruct the three-dimensional density distribution using a scale-dependent line-of-sight thickness. Tests on a simulated cloud show that such a reconstruction recovers the main features of the projected tidal eigenvalue maps \citep{Li2024a}.

The Volume Density Mapper (VDM; \citealt{Li2022,LiZhao2025,Zhao2026b}) decomposes the map into components $N_l(x,y)$ at characteristic scales $l$. Each component is assigned a Gaussian line-of-sight profile,
\begin{equation}
 n_l(x,y,z)=\frac{N_l(x,y)}{\sqrt{2\pi}\,l}\exp\left(-\frac{z^2}{2l^2}\right),\label{eq:vdm}
\end{equation}
and the reconstructed cube is $n=\sum_l n_l$. Every component is centered on the mid-plane, so the reconstructed density along each line of sight peaks at $z=0$; we call this the peak-density layer. Numerical tests of the density reconstruction yield uncertainties of 0.2--0.3 dex \citep{LiZhao2025,Zhao2026b}.

Because VDM operates on the $\NH$ map, $n$ is the molecular-hydrogen number density $n(\mathrm{H}_2)$. The mass density used in Poisson's equation is
\begin{equation}
 \rho=\mu_{\mathrm{H}_2}m_{\rm H}n(\mathrm{H}_2),\qquad \mu_{\mathrm{H}_2}=2.8,\label{eq:rho}
\end{equation}
with $m_{\rm H}=1.673\times10^{-24}$ g; the factor 2.8 includes the helium mass contribution at standard abundances. The effective density and thickness,
\begin{equation}
 \neff=\frac{\int n^2\,dz}{\int n\,dz},\qquad \leff=\frac{\NH}{\neff},\label{eq:effective}
\end{equation}
give the mass-weighted mean volume density and the effective thickness of the gas along the line of sight, respectively. For the local free-fall time we use the peak density, $n_{\rm peak}(x,y)\equiv n(x,y,0)$, with $\rho_{\rm peak}=\mu_{\mathrm{H}_2}m_{\rm H}n_{\rm peak}$.

\subsection{Tidal tensor and DC correction}\label{sec:dc}
We solve Poisson's equation on the reconstructed density cube with periodic Fourier boundary conditions, without zero-padding, and compute the tidal tensor from spectral second derivatives,
\begin{equation}
 \nabla^2\Phi=4\pi G\rho,\qquad T_{ij}=\frac{\partial^2\Phi}{\partial x_i\partial x_j}.\label{eq:poisson}
\end{equation}
In this convention positive eigenvalues denote compression and negative eigenvalues denote stretching, and Poisson's equation requires $\operatorname{Tr}(T)=\lambda_{\min}+\lambda_{\rm mid}+\lambda_{\max}=4\pi G\rho$.

In the far field of a localized mass concentration, the tidal contribution decreases as $r^{-3}$, faster than the gravitational acceleration ($r^{-2}$). For an elongated filament, the tidal field follows a shallower radial dependence at distances small compared with its length, and approaches $r^{-3}$ at larger distances (Section~\ref{sec:filament}).

Spectral second derivatives have no response at $\boldsymbol{k}=0$, so the raw tensor lacks the isotropic contribution from the mean density $\bar\rho_{\rm FFT}$ of the full FFT domain. We restore this contribution with one global constant,
\begin{equation}
 \delta_{\rm DC}=\frac{4\pi G\bar\rho_{\rm FFT}}{3},\qquad T_{ij}^{\rm corr}=T_{ij}^{\rm raw}+\delta_{\rm DC}\delta_{ij}.\label{eq:dc}
\end{equation}
This global correction restores $\operatorname{Tr}(T^{\rm corr})=4\pi G\rho$, shifts all three eigenvalues equally, and preserves the eigenvectors and eigenvalue differences. The derivation and numerical validation for the DR21 cube are given in Appendix~\ref{app:dc}.

\begin{figure*}[htbp]
 \centering\includegraphics[width=0.55\textwidth]{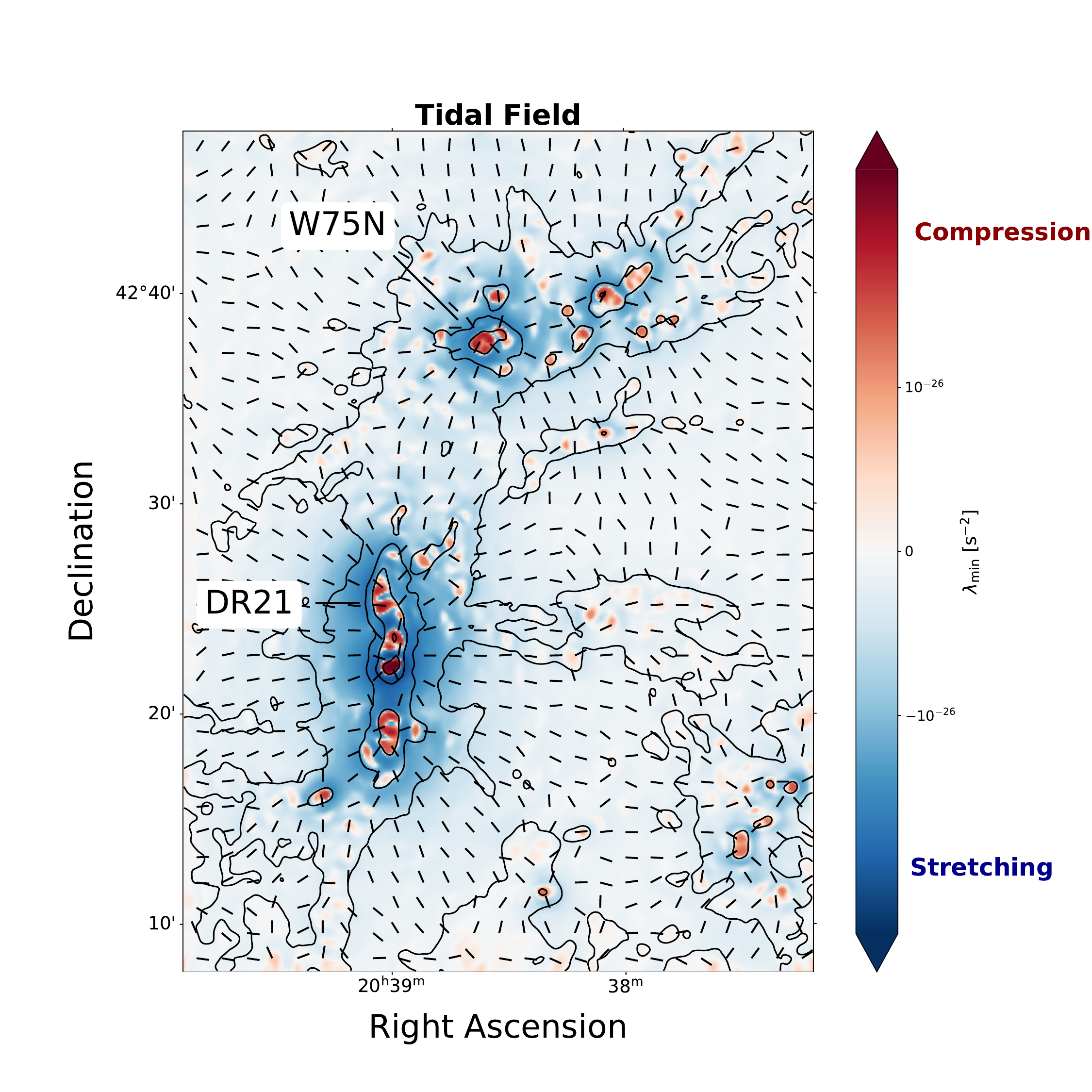}
 \caption{\textbf{Tidal topology of DR21 and W75N.} Mid-plane ($z=0$) slice of the corrected minimum tidal eigenvalue, i.e., the layer where the reconstructed line-of-sight density profile peaks. Red: $\lambda_{\min}>0$ (fully compressive); blue: $\lambda_{\min}<0$ (at least one extensive direction). Black headless segments: $\boldsymbol{e}_{\min}$. Contours: $\NH=10^{22},10^{22.5},10^{23},10^{23.5}\,\mathrm{cm}^{-2}$. DR21 has a coherent filament-centered topology; W75N shows a tidal field organized around several mass concentrations.}\label{fig:topology}
\end{figure*}

\subsection{Tidal anisotropy}\label{sec:anisotropy}
The tidal tensor describes the relative gravitational acceleration of neighboring gas elements, $\delta g_i=-T_{ij}\delta r_j$. The ordered eigenvalues $\lambda_{\min}\leq\lambda_{\rm mid}\leq\lambda_{\max}$ define the tidal topology: $\lambda_{\min}>0$ is fully compressive, whereas $\lambda_{\min}<0$ implies at least one extensive direction. The headless eigenvector $\boldsymbol{e}_{\min}$ gives the principal stretching (or weakest-compression) axis. The tidal anisotropy is
\begin{equation}
 A=\frac{\sum_{i=1}^{3}|\lambda_i-\operatorname{Tr}(T)/3|}{4\pi G\rho},\label{eq:A}
\end{equation}
which measures the strength of the deviatoric tidal field relative to the local gravitational scale $4\pi G\rho$. At fixed density, $A$ is invariant under an isotropic shift $T_{ij}\rightarrow T_{ij}+c\delta_{ij}$, including the DC correction in Equation~\eqref{eq:dc}.

We define the mass-weighted projection along the line of sight as
\begin{equation}
 \Aproj(x,y)=\frac{\int\rho A\,dz}{\int\rho\,dz}.\label{eq:Aproj}
\end{equation}
For the local quantity $A$, the reference spectrum $(-1/2,-1/2,2)$ normalized by $4\pi G\rho$ gives $\Ah=10/3$ (Appendix~\ref{app:states}). A local anisotropy of $A\geq\Ah$ implies that tidal stretching is faster than local free fall (Appendix~\ref{app:bound}). We use $\Aproj\geq\Ah$ to identify lines of sight with strong mass-weighted tidal anisotropy and to measure its spatial extent. The comparison of tidal stretching with local free fall requires the minimum eigenvalue and density at the same location; we evaluate both at the density peak ($z=0$). We use the projected anisotropy to select the strongly anisotropic subset for the timescale statistics and weight each line of sight by its column mass.

\section{Results and Discussion}\label{sec:results}
\subsection{Tidal Topology of DR21 and W75N}\label{sec:topology}
The tidal field is organized around the main ridge in DR21 and around several mass concentrations in W75N (Figure~\ref{fig:topology}). In DR21, fully compressive regions ($\lambda_{\min}>0$) are distributed along the dense ridge. On either side of the ridge, the tidal field is extensive ($\lambda_{\min}<0$), and the projected principal stretching axes $\boldsymbol{e}_{\min}$ tend to be oriented across the ridge. The dense ridge contains several local mass concentrations, yet the stretching axes in the surrounding gas are coherent over much larger scales. The gravity of the ridge as a whole thus appears to govern the orientation of the surrounding tidal field.

In W75N, compressive regions are associated with several concentrations of dense gas. The stretching axes change in orientation around these concentrations, rather than following a common pattern across the region. The contrast in the tidal field reflects the distribution of mass: DR21 contains a prominent ridge, whereas W75N contains several mass concentrations, including filamentary structures \citep{Cao2019,LiQiu2026}. In DR21, several regions of gravitational compression share a common large-scale tidal field; in W75N, the tidal field is organized around several mass concentrations. We describe these gravitational architectures as centralized and multipolar, respectively.

The stellar morphology also differs between the two regions. \citet{Kumar2007} found the DR21/DR21(OH) cluster to be elongated and substructured along the dense filament, whereas the W75N cluster is more circular and centrally concentrated. The stellar--filament alignment in DR21 is consistent with the importance of the ridge on large scales.

\begin{figure*}[htbp]
 \centering\includegraphics[width=0.92\textwidth]{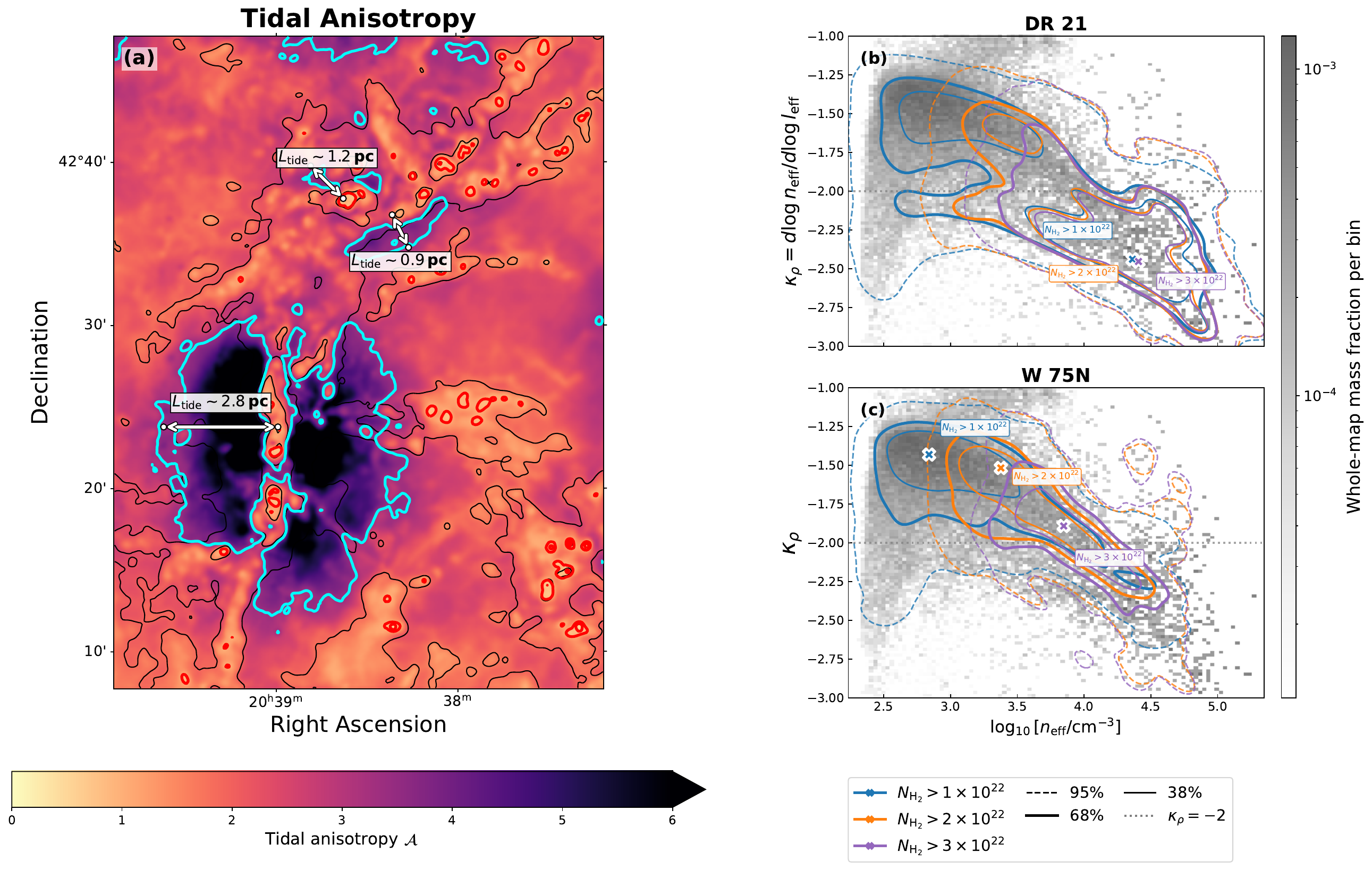}
 \caption{\textbf{Tidal anisotropy, representative high-anisotropy extents, and density--scale structure.} (a) Mass-weighted projected tidal anisotropy $\Aproj=\int\rho A\,dz/\int\rho\,dz$. Black contours: column density. Red contours: $\Aproj=2/3,1$. Cyan: $\Aproj=10/3$. White double-headed arrows indicate representative $\Lt$ measurements from a dominant filament or clump structure to the outer contour of the associated region of high projected anisotropy. The DR21 ridge has $\Lt\sim2.8$ pc, while two regions in W75N have representative values of $\Lt\sim1.2$ and 0.9 pc. (b,c) Column-density-weighted $(\log\neff,\kappa_\rho)$ distributions for DR21 and W75N. Contours at three $\NH$ thresholds enclose 38\%, 68\%, and 95\% of the weighted mass; crosses mark the KDE peaks. Dotted line: $\kappa_\rho=-2$. The DR21 peak is nearly stationary; the W75N peak shifts with threshold. The same projected threshold $\Aproj\geq\Ah$ is used to select the strongly anisotropic subset in Figure~\ref{fig:timescale}.}\label{fig:anisotropy}
\end{figure*}

\subsection{Tidal Anisotropy and Density--Scale Structure}\label{sec:anisotropy_results}
Strong tidal anisotropy extends much farther from the main ridge in DR21 than from individual mass concentrations in W75N (Figure~\ref{fig:anisotropy}a). In DR21, the connected region of high projected anisotropy surrounds the ridge and extends into gas at substantially lower column densities. In W75N, several smaller regions of high projected anisotropy are associated with local mass concentrations, consistent with the changes in orientation of the stretching axes in Figure~\ref{fig:topology}.

We measure a representative projected extent $\Lt$ from a filament spine or local clump center to the outer $\Aproj=\Ah$ contour of the associated high-anisotropy region. The DR21 ridge has $\Lt\sim2.8$ pc, whereas two regions in W75N have representative extents of $\sim0.9$--1.2 pc. Since $A$ measures the deviatoric tidal strength relative to $4\pi G\rho$, this extent depends on both the tidal field and the density of the surrounding gas. The length $\Lt$ is a measure of the extent of strong tidal anisotropy, rather than a gravitational binding radius.

The mass distribution that produces the tidal field can also be characterized by its density--scale relation. A massive ridge can contribute gas over a wide range of column densities. Where several mass concentrations coexist, the relative contributions of structures with different density--scale relations may change as the column-density threshold increases. The persistence of a characteristic density and density exponent across these thresholds can therefore help us understand how the mass is concentrated. The density exponent $\kappa_\rho=d\log\rho/d\log r$ describes how rapidly the density increases toward smaller scales \citep{Shu1977,Li2018}. Measurements show that the exponent varies within molecular clouds and even along individual filaments \citep{LiZhou2022,Zhao2026a}. We estimate the local exponent $\kappa_\rho=d\log\neff/d\log\leff$ using Adjacent Correlation Analysis (ACA; \citealt{Li2025a,Li2025b}; Appendix~\ref{app:aca} and \citealt{Zhao2026a}).

Since $\NH=\neff\leff$, a high column density can arise from a high volume density or a large thickness. To examine the gas selected at different column densities, we construct column-density-weighted kernel density estimates (KDEs) in the $(\log_{10}\neff,\kappa_\rho)$ plane at three column-density thresholds (Figure~\ref{fig:anisotropy}b,c). The KDE peak identifies the characteristic effective density and density exponent around which the gas mass is most concentrated. Its displacement $\Delta_{\rm KDE}$ between the lowest and highest column-density thresholds measures how much this characteristic density structure changes (Appendix~\ref{app:kde}).

In DR21, gas at relatively high volume density and with a steep density--scale relation is already prominent at the lowest threshold. The KDE peak changes little as the threshold increases from $10^{22}$ to $3\times10^{22}\,\mathrm{cm}^{-2}$, with $\Delta_{\rm KDE}=0.10$. This is consistent with the dense ridge remaining prominent over this range of column densities. In W75N, the peak at the lowest threshold lies at lower density and has a less negative density exponent. As the threshold increases, the peak moves toward higher $\neff$ and more negative $\kappa_\rho$, with $\Delta_{\rm KDE}=1.00$. The characteristic density--scale relation of the selected gas in W75N therefore changes with column density, consistent with the coexistence of structures with different density properties \citep{Zhao2026a}. Spatial maps of $\neff$, $\leff$, and $\kappa_\rho$ are shown in Appendix~\ref{app:density}.

The DR21 ridge is prominent in both the density structure and the tidal field. Its tidal influence, however, extends several parsecs from the dense gas. We examine the origin of this large extent using the gravitational field of a filament.

\begin{figure*}[htbp]
 \centering
 % Replace this PNG with the original vector PDF when available.
 \IfFileExists{Figure_filament_tidal_model_final.pdf}{\includegraphics[width=0.93\textwidth]{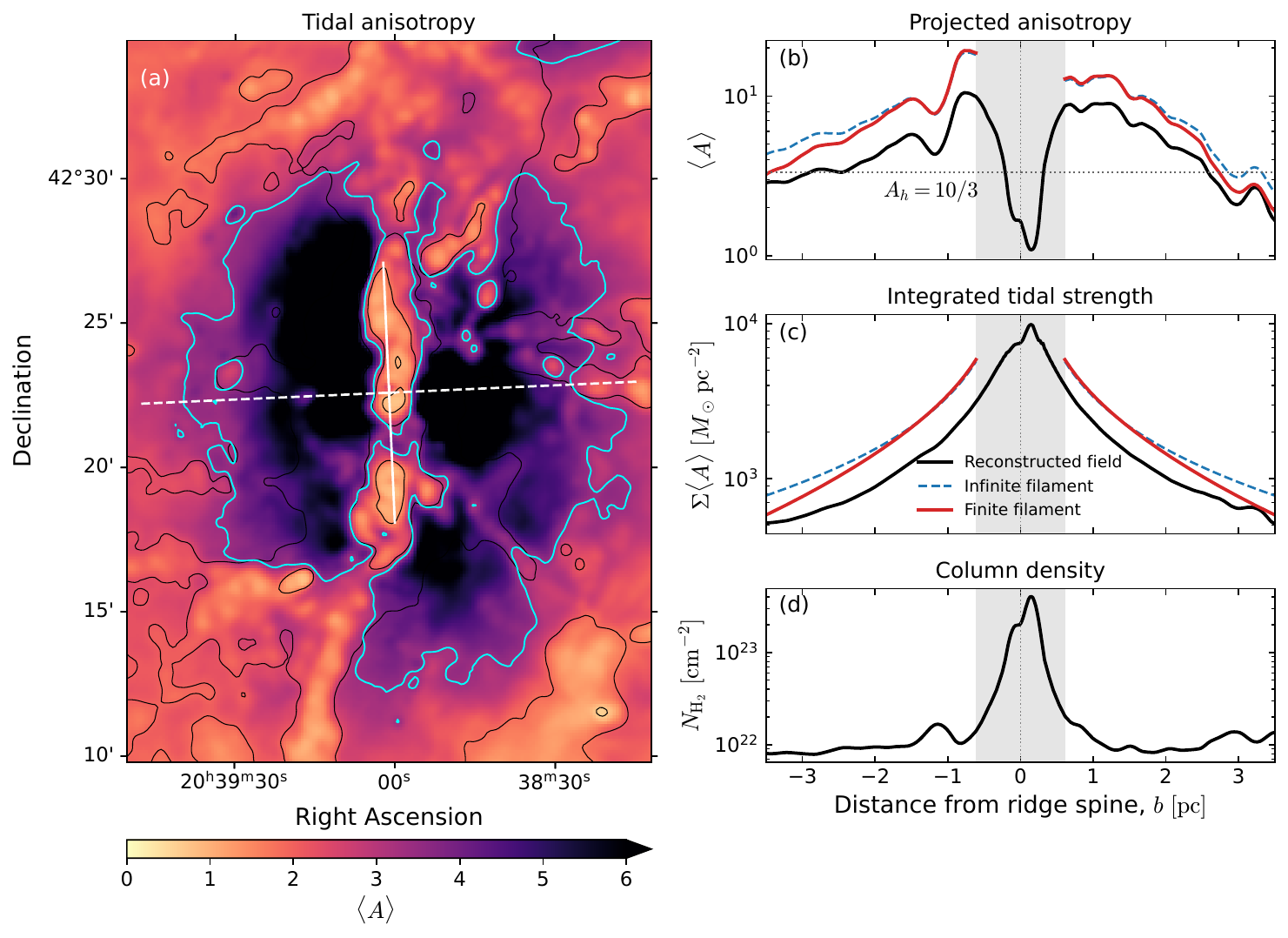}}{\includegraphics[width=0.93\textwidth]{filament_tidal_model.png}}
 \caption{\textbf{Tidal field across the DR21 ridge.} (a) Mass-weighted projected tidal anisotropy $\Aproj$ (Equation~\ref{eq:Aproj}). The solid white segment marks the adopted ridge axis, the cross its midpoint, and the dashed white line the transverse cut. Black contours: $\NH=10^{22},10^{22.5},10^{23},10^{23.5}\,\mathrm{cm}^{-2}$. Cyan contour: $\Aproj=10/3$. (b) Projected tidal anisotropy along the cut. Black: reconstructed field; dashed blue and solid red: infinite- and finite-filament models, respectively, with the line mass and length from \citet{Hennemann2012}. (c) The product $\Sigma\Aproj$, proportional to the integrated deviatoric tidal strength (Equation~\ref{eq:integrated}). (d) Column density along the same cut. The shaded interval $|b|<0.6$ pc is excluded from the line-mass models.}\label{fig:filament}
\end{figure*}

\subsection{The Gravitational Extent of the DR21 Ridge}\label{sec:filament}
A filament attracts the surrounding gas while stretching it in the radial direction. Outside an infinitely long filament of line mass $\mu$, the gravitational acceleration at a distance $R$ from its axis is $g_R=-2G\mu/R$ \citep{Heitsch2013}. In the radial, azimuthal, and longitudinal directions, the eigenvalues of the filament's tidal tensor are
\begin{equation}
 (\lambda_R^{\rm fil},\lambda_\phi^{\rm fil},\lambda_\parallel^{\rm fil})=\frac{2G\mu}{R^2}(-1,1,0).\label{eq:infinite}
\end{equation}
Although the radial acceleration is inward, its magnitude increases toward the filament, so the relative acceleration of radially separated gas elements is extensive. This extensive tide opposes radial contraction of the surrounding gas, while the azimuthal tide is compressive \citep{Jog2013,Li2024a,Li2024b}.

The strength of the filament's tide relative to local gravity depends on its line mass and the density of its surroundings. If the filament dominates the deviatoric field and the tidal contribution of the ambient gas is approximately isotropic, Equations~\eqref{eq:A} and \eqref{eq:infinite} give
\begin{equation}
 A(R)\simeq\frac{\mu}{\pi R^2\rho_{\rm env}(R)}.\label{eq:filA}
\end{equation}
Here $\rho_{\rm env}$ is the local mass density of the surrounding gas. The quantity $\mu/(\pi R^2)$ is the mean density obtained by distributing the filament line mass over a cylinder of radius $R$. The radius at which $A=\Ah$ therefore satisfies $R_{\rm h}^2\rho_{\rm env}(R_{\rm h})\simeq\mu/(\pi\Ah)$. For an approximately uniform ambient density,
\begin{equation}
 R_{\rm h}\simeq\left(\frac{\mu}{\pi\Ah\rho_{\rm env}}\right)^{1/2}.\label{eq:filradius}
\end{equation}
A filament of greater line mass can influence the conditions for local contraction over a larger radial extent at fixed ambient density. In a nonuniform medium this extent is also set by the decline of the ambient density. In the same idealized model the total radial eigenvalue is $\lambda_R=4\pi G\rho_{\rm env}(1/3-A/2)$, so that the stretching term dominates the local isotropic compression at $A>2/3$.

The finite length of a filament changes the radial decline of its tidal field. For a straight filament of uniform line mass and length $L=2a$, in the plane perpendicular to its axis through its midpoint, the eigenvalues outside the line mass are $(-K_1-K_2,K_1,K_2)$, where
\begin{equation}
 K_1=\frac{2G\mu a}{R^2\sqrt{R^2+a^2}},\qquad K_2=\frac{2G\mu a}{(R^2+a^2)^{3/2}}.\label{eq:finite}
\end{equation}
The radial tide approaches that of the infinite filament for $R\ll a$ and that of a point mass $M=2a\mu$ for $R\gg a$; its magnitude changes from an $R^{-2}$ to an $R^{-3}$ dependence.

The projected anisotropy follows from integrating the deviatoric tidal strength along the line of sight. Let $\Sigma=\int\rho\,dz$ and $b$ denote the signed projected distance from the ridge spine. Then Equation~\eqref{eq:Aproj} gives
\begin{equation}
 \Sigma\Aproj=\frac{1}{4\pi G}\int\sum_i\left|\lambda_i-\frac{\operatorname{Tr}(T)}{3}\right|\,dz.\label{eq:integrated}
\end{equation}
For a filament in the plane of the sky, $R^2=b^2+z^2$. With the same assumptions as in Equation~\eqref{eq:filA}, the infinite and finite filament models yield, respectively,
\begin{align}
 \Sigma\Aproj_{\rm inf}(b)&=\frac{\mu}{\pi}\int_{-Z}^{Z}\frac{dz}{R^2},\label{eq:projinf}\\
 \Sigma\Aproj_{\rm fin}(b)&=\frac{\mu a}{\pi}\int_{-Z}^{Z}\left[\frac{1}{R^2\sqrt{R^2+a^2}}+\frac{1}{(R^2+a^2)^{3/2}}\right]dz,\label{eq:projfin}
\end{align}
where $Z$ is the half-depth of the integration. For sufficiently large $Z$, the infinite-filament expression approaches $\mu/|b|$, and the projected high-anisotropy boundary satisfies $|b_{\rm h}|\Sigma(b_{\rm h})\simeq\mu/\Ah$. The finite-filament expression approaches the same dependence at $|b|\ll a$, and scales as $|b|^{-2}$ in the limit $a\ll |b|\ll Z$.

The analytical filament models account for the order of magnitude and radial decline of the tidal field around the DR21 ridge (Figure~\ref{fig:filament}). We adopt $\mu=3670\,\Ms\,\mathrm{pc}^{-1}$ and $L=4.1$ pc from the Herschel measurements of \citet[Table 1]{Hennemann2012}. The profiles are evaluated along a transverse cut near the middle of the ridge, with the same finite line-of-sight integration range as the reconstructed field. We normalize the analytical profiles by the observed surface density $\Sigma=\mu_{\mathrm{H}_2}m_{\rm H}\NH$ along the cut. No parameters are fitted to the reconstructed tidal profiles. The finite-filament model gives values of $\Aproj$ larger than the reconstructed values by median factors of 1.63 and 1.42 on the negative and positive sides of the cut, respectively, over $0.6<|b|\leq3.5$ pc. The comparison of $\Sigma\Aproj$ shows a similar radial decline in the integrated deviatoric tidal strength, without the common surface-density normalization. Toward the ends of the cut, accounting for the finite length generally brings the model closer to the reconstructed profile. The models neglect the transverse mass distribution of the ridge and the tides of neighboring structures. The line mass measured by \citet{Hennemann2012} includes gas within a radius of 1.15 pc, whereas the central width is 0.34 pc. Concentrating this mass on the axis can contribute to the excess at small $|b|$.

The tidal influence of the DR21 ridge extends well beyond the region occupied by its dense gas. Along the transverse cut, the column density approaches the ambient level at distances of order 1 pc, whereas strong tidal anisotropy extends to distances of several parsecs. A filament can thus be characterized by both its geometrical width and the extent of its tidal influence.

The smaller extents of strong projected tidal anisotropy around individual structures in W75N are consistent with the dependence of tidal extent on mass distribution, geometry, and ambient density.

\begin{figure*}[htbp]
 \centering\includegraphics[width=0.86\textwidth]{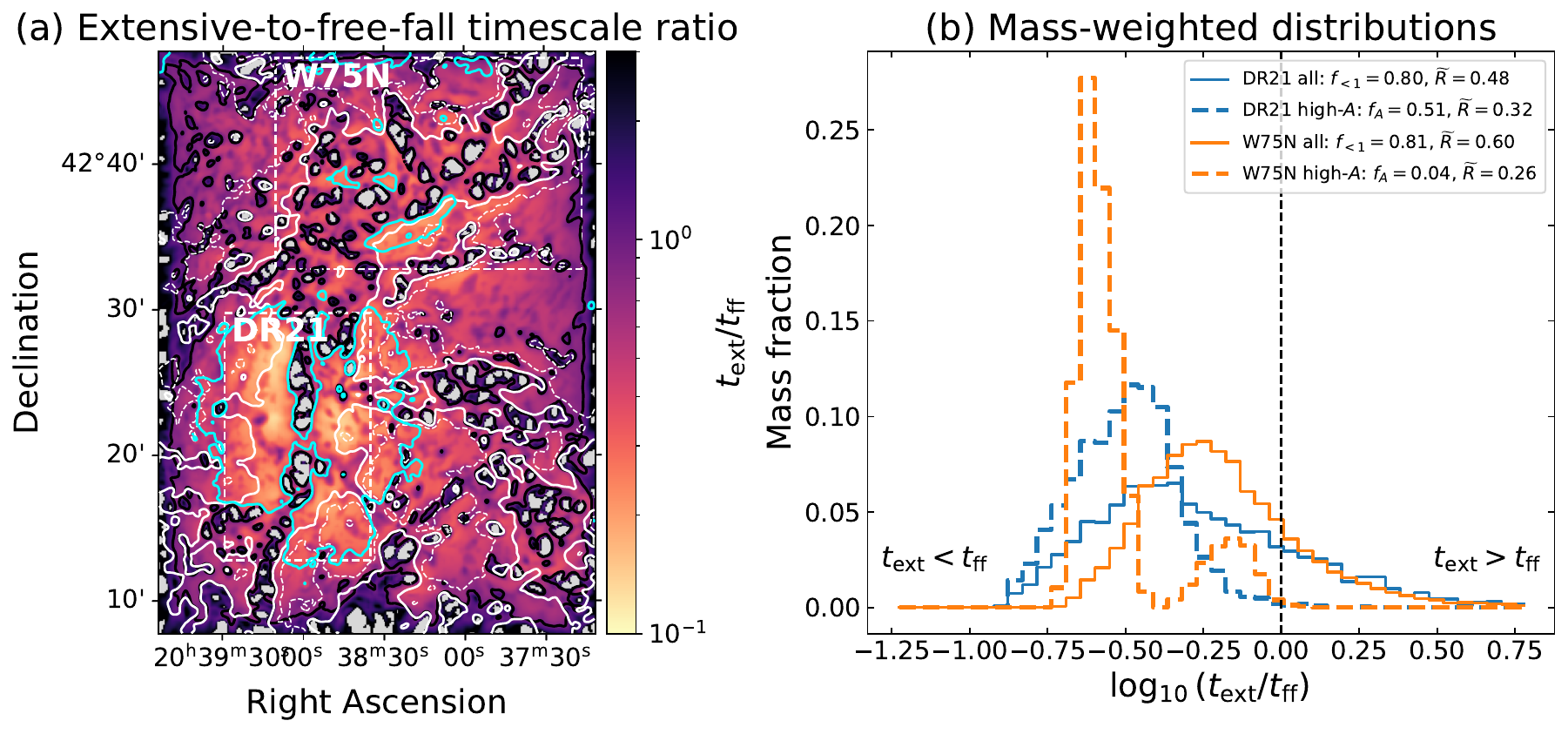}
 \caption{\textbf{Local tidal-to-free-fall timescale ratio at the density peak.} (a) Map of $\Rt=\texten/\tff$ in the density-peak layer ($z=0$), i.e., the single line-of-sight layer where the reconstructed Gaussian density profile reaches its maximum. Both $\rho_{\rm peak}$ and $\lambda_{\min}$ are evaluated in this layer. Shown only where $\lambda_{\min}<0$. Black contour: $\Rt=1$. White dashed and solid contours: $\NH=8\times10^{21}$ and $10^{22}\,\mathrm{cm}^{-2}$, respectively. Cyan contour: $\Aproj=10/3$. Dashed boxes: DR21 and W75N regional masks. (b) Mass-weighted distributions of $\log_{10}\Rt$ for DR21 and W75N. Solid: all lines of sight with $\lambda_{\min}<0$ at $z=0$; dashed: the subset with $\Aproj\geq10/3$. Vertical dashed line: $\Rt=1$. Weights are pixel masses, $m_p\propto\NH\times\text{pixel area}$. Each histogram is normalized separately to unit summed mass fraction. Legend values: $f_{<1}$ and the mass-weighted median $\widetilde{\Rt}$ for the full extensive sample; $f_A$ (fraction of its column mass with $\Aproj\geq\Ah$) and $\widetilde{\Rt}$ for that subset.}\label{fig:timescale}
\end{figure*}

\subsection{Tidal-to-Free-Fall Timescale Ratios}\label{sec:timescale}
The minimum tidal eigenvalue sets a characteristic time for gravitational stretching. Where $\lambda_{\min}<0$, we compare this time with the local free-fall time,
\begin{equation}
 \texten=(-\lambda_{\min})^{-1/2},\qquad \tff=\left(\frac{3\pi}{32G\rho_{\rm peak}}\right)^{1/2},\label{eq:times}
\end{equation}
and define $\Rt=\texten/\tff$. Both $\rho_{\rm peak}$ and $\lambda_{\min}$ are evaluated at $z=0$. The statistics include only lines of sight with $\lambda_{\min}<0$ at the density peak, within the regional masks shown in Figure~\ref{fig:timescale}.

Within each region we define the column-mass fractions with $\Rt<1$ and $\Aproj\geq10/3$, and the mass-weighted median $\widetilde{\Rt}$:
\begin{align}
 f_{<1}&=\frac{M(\lambda_{\min}<0,\Rt<1)}{M(\lambda_{\min}<0)},\nonumber\\
 f_A&=\frac{M(\lambda_{\min}<0,\Aproj\geq10/3)}{M(\lambda_{\min}<0)},\label{eq:fractions}\\
 \widetilde{\Rt}&=\operatorname{wmed}_{p}(\mathcal{R}_{{\rm tide},p};m_p),\label{eq:medianR}
\end{align}
with $m_p\propto\NH(x_p,y_p)\times(\text{pixel area})$. Here $\operatorname{wmed}$ denotes the weighted median. We calculate $\widetilde{\Rt}$ separately for all lines of sight classified as extensive at $z=0$ and for those with $\Aproj\geq\Ah$ within this sample.

The tidal stretching time is shorter than the local free-fall time for about 80\% of the column mass in lines of sight subject to extensive tides at the density peak in either region. Figure~\ref{fig:timescale} gives
\begin{equation}
 f_{<1}=0.80\ ({\rm DR21}),\qquad f_{<1}=0.81\ ({\rm W75N}),\label{eq:f1}
\end{equation}
with mass-weighted medians $\widetilde{\Rt}=0.48$ and 0.60, respectively.

Among lines of sight classified as extensive at the density peak, a larger fraction of the column mass has strong projected tidal anisotropy in DR21,
\begin{equation}
 f_A=0.51\ ({\rm DR21}),\qquad f_A=0.04\ ({\rm W75N}).\label{eq:fA}
\end{equation}
For the subset with $\Aproj\geq\Ah$, the mass-weighted medians $\widetilde{\Rt}$ are 0.32 in DR21 and 0.26 in W75N. Although this subset contains a much smaller fraction of the extensive column mass in W75N, the stretching times at its density peaks are also short compared with local free fall. A local anisotropy of $A\geq10/3$ implies $\Rt\leq4/(\sqrt{3}\pi)\simeq0.735$ at the same position (Appendix~\ref{app:bound}), but the projected threshold does not impose this bound at $z=0$. In DR21 the lines of sight with strong projected anisotropy surround the ridge; in W75N they are associated with several mass concentrations. The two regions differ more in the spatial distribution and column-mass fraction of strong tidal anisotropy than in the fraction with short tidal timescales.

\begin{figure*}[htbp]
 \centering\includegraphics[width=0.86\textwidth]{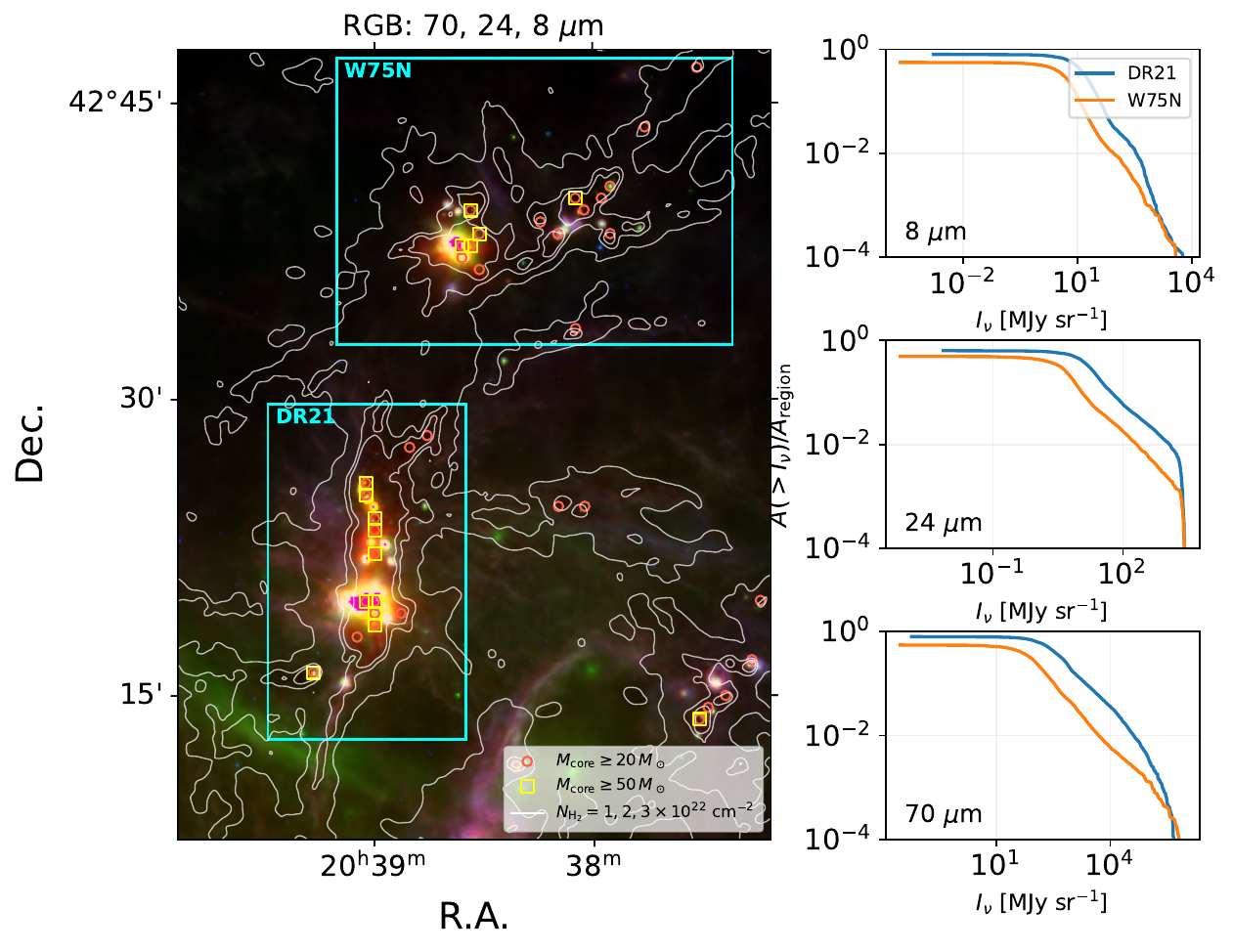}
 \caption{\textbf{Massive cores and infrared emission in DR21 and W75N.}
    Left: Three-color infrared composite with 70, 24, and 8~$\mu$m emission shown in red, green, and blue, respectively, with column-density contours and the DR21 and W75N regional masks.
    Core positions are taken from \citet{LiQiu2026}; markers highlight $M_{\rm core}\geq20\,M_\odot$ and $M_{\rm core}\geq50\,M_\odot$.
    Right: Complementary cumulative area fractions $A(>I_\nu)/A_{\rm region}$ at 8, 24, and 70~$\mu$m after local-background subtraction.
    DR21 has a larger fraction of its area at high infrared surface brightness, particularly at 24 and 70~$\mu$m.}
    \label{fig:cores}
\end{figure*}
\begin{deluxetable*}{lccc}
\tablecaption{Summary of tidal, gas, and core measurements\label{tab:summary}}
\tablehead{\colhead{Quantity}&\colhead{DR21}&\colhead{W75N}&\colhead{Unit / definition}}
\startdata
$f_A$ &0.51&0.04&fraction with $\Aproj\geq\Ah$\\
$\Lt$ &$\sim2.8$&$\sim0.9$--1.2&pc\\
$M_{\rm gas}$ &$1.54\times10^4$&$1.46\times10^4$&$\Ms$\\
$f_{\rm high-N}$ &0.567&0.199&$M(\NH>3\times10^{22})/M_{\rm gas}$\\
$f_{<1}$ &0.80&0.81&fraction with $\Rt<1$\\
$\widetilde{\Rt}$ &0.48&0.60&mass-weighted median\\
$M_{\rm core,max}$ &507&147&$\Ms$\\
$f_{\rm core,20}$ &0.108&0.0584&$M(>20\,\Ms)/M_{\rm gas}$\\
$f_{\rm core,50}$ &0.0988&0.0355&$M(>50\,\Ms)/M_{\rm gas}$\\
$M_{\rm core,>50}/M_{\rm high-N}$ &0.174&0.178&mass reservoir ratio
\enddata
\tablecomments{$\Lt$ is a representative extent of strong projected tidal anisotropy (Figure~\ref{fig:anisotropy}a). The fractions $f_A$ and $f_{<1}$ are weighted by column mass; $\lambda_{\min}$ and $\Rt$ are evaluated at $z=0$, and the anisotropy selection uses $\Aproj$. $f_A=M(\lambda_{\min}<0,\Aproj\geq10/3)/M(\lambda_{\min}<0)$ and $f_{<1}=M(\lambda_{\min}<0,\Rt<1)/M(\lambda_{\min}<0)$. $M_{\rm high-N}=M(\NH>3\times10^{22}\,\mathrm{cm}^{-2})$; the mass reservoir ratio is $M(M_{\rm core}>50\,\Ms)/M_{\rm high-N}$. Core quantities are measured from the Cygnus-X core catalog \citep{LiQiu2026} within the same regional masks as the tidal analysis.}
\end{deluxetable*}
\subsection{High-Column-Density Gas, Massive Cores, and Infrared Morphology}\label{sec:cores}
DR21 contains a larger fraction of its gas at high column density and in high-mass cores. The two regional masks contain nearly the same total gas mass, $M_{\rm gas}=1.54\times10^4\,\Ms$ in DR21 and $1.46\times10^4\,\Ms$ in W75N. Above $\NH=3\times10^{22}\,\mathrm{cm}^{-2}$, the high-column-density mass fractions are $f_{\rm high-N}=0.567$ and 0.199, respectively, consistent with the cloud-dependent column-density results across Cygnus-X reported by \citet{XingQiu2026}. Core properties are taken from the uniform Cygnus-X catalog of \citet{LiQiu2026}; we use $20\,\Ms$ as the high-mass threshold adopted in that work and quote the $>50\,\Ms$ subset for the most massive cores. The most massive cores are $507\,\Ms$ in DR21 and $147\,\Ms$ in W75N. Cores above $20\,\Ms$ contain 10.8\% and 5.84\% of the regional gas mass, and cores above $50\,\Ms$ contain 9.88\% and 3.55\%, respectively. The larger high-mass-core fraction in DR21 is consistent with the Cygnus-X-wide association of massive cores with filamentary gas \citep{LiQiu2026}.

The differences between DR21 and W75N are also seen in their infrared emission (Figure~\ref{fig:cores}).
DR21 has a larger fraction of its area at high infrared surface brightness, particularly at 24 and 70~$\mu$m.
The integrated fluxes in DR21 exceed those in W75N by factors of 2.16, 2.40, and 2.00 at 8, 24, and 70~$\mu$m, respectively.
The corresponding ratios of the 99th-percentile intensities are 3.86, 5.85, and 8.41.
The larger mass fraction in massive cores in DR21 is accompanied by stronger infrared emission.
The infrared measurements are described in Appendix~\ref{app:ir}.

Mass assembly onto the DR21 ridge may compete with fragmentation in its surroundings. As gas is attracted toward the ridge, radial tidal stretching can hinder independent contraction of the surrounding gas \citep{Li2024a,Li2024b}. The observed accretion flows toward the ridge \citep{Schneider2010,Hennemann2012,Hu2021} are consistent with this picture, and tidal effects on fragmentation have also been discussed for the fibers in DR21(OH) \citep{Cao2022}. The formation of fragments can thus be related to mass transport on larger scales, as discussed by \citet{Li2024c} and \citet{LiZhaoLu2025}. In W75N, mass assembly takes place in a tidal field organized around several mass concentrations.

The larger massive-core mass in DR21 accompanies a larger reservoir of high-column-density gas. The ratios of the total mass in cores above $50\,\Ms$ to the mass at $\NH>3\times10^{22}\,\mathrm{cm}^{-2}$ are similar, 0.174 and 0.178 in DR21 and W75N, respectively (Table~\ref{tab:summary}). On clump scales, \citet{Jiao2025} likewise report a higher star formation rate (SFR) and a larger self-gravitating gas mass in DR21 S than in W75N, while the SFR per unit self-gravitating gas mass is similar ($\simeq4.3\times10^{-9}$ and $4.8\times10^{-9}\,\mathrm{yr}^{-1}$).

\section{Conclusions}\label{sec:conclusions}
The tidal field of a cloud can be organized by one dominant mass concentration or by several local structures. In DR21, several regions of gravitational compression lie along the main ridge, while the stretching axes in the surrounding gas are coherent on larger scales. In W75N, the stretching axes are associated with several mass concentrations. We describe these tidal fields as centralized and multipolar, respectively. The characteristic effective density and density exponent traced by the KDE peak remain nearly unchanged in DR21 as the column-density threshold increases ($\Delta_{\rm KDE}=0.10$), but shift toward higher densities and more negative exponents in W75N ($\Delta_{\rm KDE}=1.00$). These differences are consistent with the prominence of a single ridge in DR21 and the coexistence of several structures in W75N.

The tidal influence of a massive filament can extend well beyond its dense gas. The projected extent of strong anisotropy is $\Lt\sim2.8$ pc around the DR21 ridge, compared with representative extents of $\sim0.9$--1.2 pc around mass concentrations in W75N. Analytical filament models using the observed line mass and length of the DR21 ridge account for the order of magnitude and radial decline of the surrounding tidal field. In the idealized cylindrical limit, the anisotropy is given by the ratio of the mean density of the filament mass within a cylinder of radius $R$ to the local ambient density. The extent of strong tidal anisotropy therefore depends on both the mass distribution of the filament and the density of its surroundings.

Regions with similar relative tidal timescales can differ in the spatial organization of their tidal fields and in the concentration of mass. In both DR21 and W75N, tidal stretching is faster than local free fall for about 80\% of the column mass classified as extensive at the density peak. Yet the fractions of this column mass with strong projected tidal anisotropy ($\Aproj\geq\Ah$) are 0.51 and 0.04, respectively. At nearly equal total gas masses, DR21 contains more high-column-density gas and more mass in massive cores than W75N. The similar ratios of mass in cores above $50\,\Ms$ to high-column-density gas mass indicate that the difference in massive-core mass accompanies a difference in the amount of dense gas. By attracting gas while opposing its independent contraction along the stretching direction, a dominant mass concentration may favor the assembly of gas on larger scales.

The local density sets the free-fall time, but it does not determine the directions or timescales of gravitational stretching. These depend on the distribution of mass throughout the cloud. The tidal tensor is therefore needed to describe the gravitational deformation of gas in a collapsing cloud. Comparing the stretching axes and tidal timescales with gas motions may help reveal where gas undergoes local contraction and where it is assembled into larger structures.

% \begin{acknowledgments}
% We used DeepSeek to assist with language polishing and improving the clarity of the manuscript text. All scientific content, analyses, and final wording were reviewed and verified by the authors, who take full responsibility for the manuscript.
% \end{acknowledgments}

% Bibliography: 46 entries.

\appendix
\renewcommand{\theHequation}{app.\thesection.\arabic{equation}}
\section{Method Details}\label{app:methods}
\subsection{VDM Reconstruction}\label{app:vdm}
The column-density map is decomposed into components $N_l(x,y)$ at characteristic scales $l$, each centered on the mid-plane and assigned a Gaussian line-of-sight profile:
\begin{equation}
 n_l(x,y,z)=\frac{N_l(x,y)}{\sqrt{2\pi}\,l}\exp\left(-\frac{z^2}{2l^2}\right).
\end{equation}
The cube is $n=\sum_l n_l$. The peak-density layer ($z=0$) and the effective density and thickness are defined in Section~\ref{sec:vdm}.

\subsection{Global DC Correction and Numerical Validation}\label{app:dc}
The missing zero-frequency contribution is restored by a single global isotropic offset determined by the mean density of the FFT domain. For each nonzero Fourier mode,
\begin{equation}
 \hat\Phi(\boldsymbol{k})=-\frac{4\pi G\hat\rho(\boldsymbol{k})}{k^2},\qquad k^2=k_x^2+k_y^2+k_z^2,
\end{equation}
and the Hessian convention used in this work gives
\begin{equation}
 \hat T_{ij}=-k_i k_j\hat\Phi.
\end{equation}
The factor $k_i k_j$ removes the zero-frequency mode from every spectral second derivative. The raw Fourier solution therefore responds to the density contrast $\rho-\bar\rho_{\rm FFT}$ rather than to $\rho$ itself, so
\begin{equation}
 \operatorname{Tr}(T^{\rm raw})=4\pi G(\rho-\bar\rho_{\rm FFT}).
\end{equation}
The missing trace term is spatially constant, and we restore it with an isotropic correction,
\begin{equation}
 \delta_{\rm DC}=\frac{4\pi G\bar\rho_{\rm FFT}}{3},\qquad T_{ij}^{\rm corr}=T_{ij}^{\rm raw}+\delta_{\rm DC}\delta_{ij}.
\end{equation}
This restores $\operatorname{Tr}(T^{\rm corr})=4\pi G\rho$. Because the correction is proportional to the identity matrix, it shifts all three eigenvalues by the same amount and leaves the eigenvectors and eigenvalue differences unchanged.

\begin{figure*}[htbp]
 \centering\includegraphics[width=0.91\textwidth]{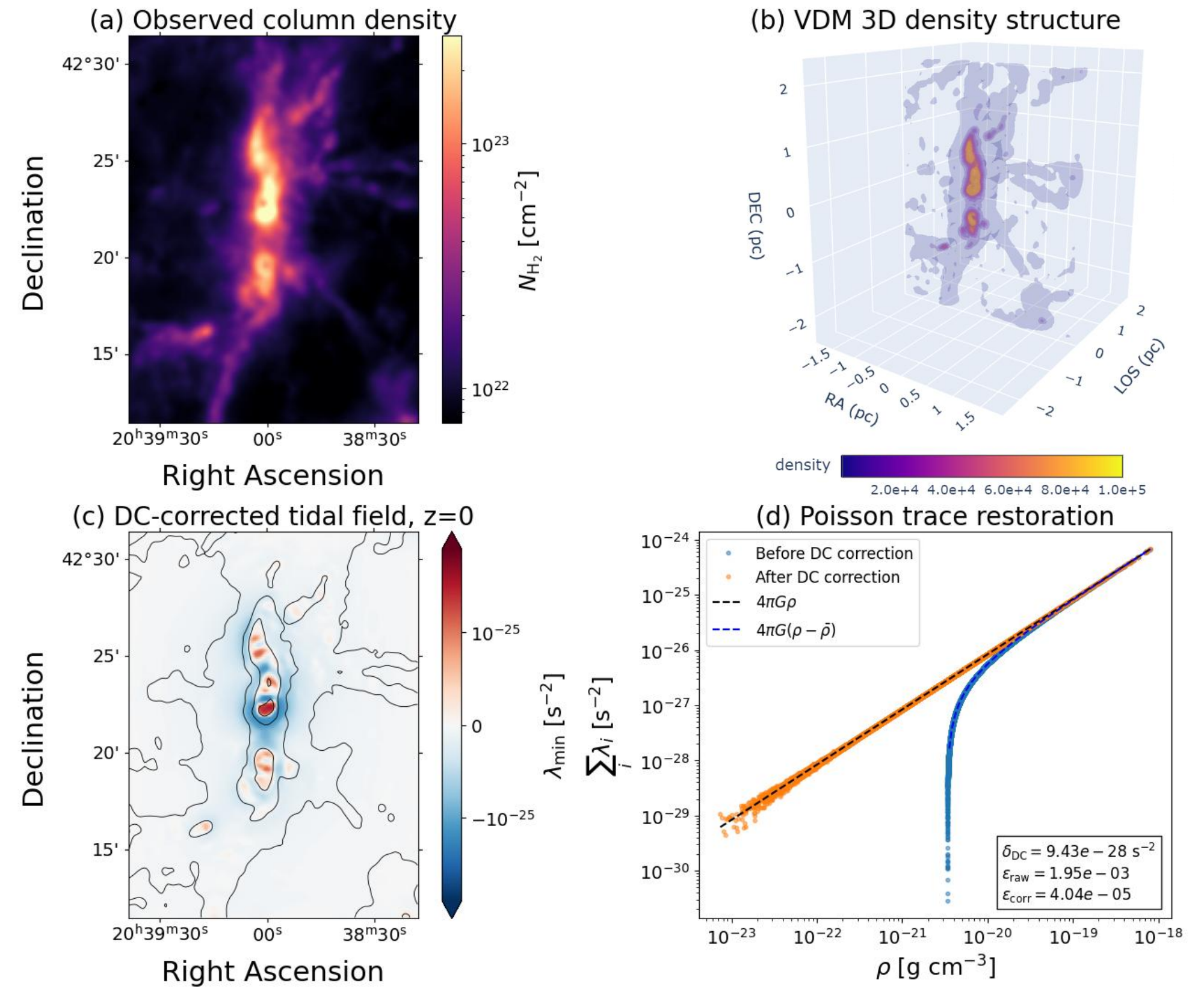}
 \caption{\textbf{Global DC correction for the DR21 tidal calculation.} (a) Observed $\NH$ column density. (b) VDM-reconstructed three-dimensional $n(\mathrm{H}_2)$ density structure; the color scale is in $\mathrm{cm}^{-3}$. (c) DC-corrected $\lambda_{\min}$ at the density peak ($z=0$). (d) Sum of the tidal eigenvalues before and after the global correction. The raw trace follows $4\pi G(\rho-\bar\rho_{\rm FFT})$, whereas the corrected trace follows the Poisson relation $4\pi G\rho$. The same offset, $\delta_{\rm DC}=9.43\times10^{-28}\,\mathrm{s}^{-2}$, is applied to every voxel; $\epsilon_{\rm tr}$ decreases from $1.95\times10^{-3}$ to $4.04\times10^{-5}$. Negative raw traces are omitted from panel (d).}\label{fig:dc}
\end{figure*}

Figure~\ref{fig:dc} verifies the correction directly for the DR21 test cube. Before the correction, the trace follows $4\pi G(\rho-\bar\rho_{\rm FFT})$; after the same global offset is added to every voxel, it follows $4\pi G\rho$ over the full density range. We quantify the trace residual as
\begin{equation}
 \epsilon_{\rm tr}=\frac{\max|\sum_i\lambda_i-4\pi G\rho|}{\max|4\pi G\rho|}.
\end{equation}
For DR21, $\delta_{\rm DC}=9.43\times10^{-28}\,\mathrm{s}^{-2}$ and $\epsilon_{\rm tr}$ decreases from $1.95\times10^{-3}$ before correction to $4.04\times10^{-5}$ after correction. The global shift can change the sign of eigenvalues close to zero and therefore affects the compressive/extensive classification and $\Rt$, whereas the anisotropy $A$ is unchanged because it depends on eigenvalue differences from the isotropic part. The correction restores only the missing isotropic $k=0$ contribution; the nonzero Fourier modes, including the deviatoric tidal field and eigenvectors, retain the periodic boundary condition of the Fourier solver.

\noindent\textit{Code availability.} The tidal calculation, including CPU and GPU implementations of the Fourier solver and the global DC correction, is available in the GravTide3D package at \url{https://github.com/meng-ke/GravTide3D}.

\subsection{ACA Density Exponent}\label{app:aca}
At each position, we set $p_1=\log\leff$ and $p_2=\log\neff$, so the local slope in the $(\log\leff,\log\neff)$ plane is the density exponent $\kappa_\rho=dp_2/dp_1$. The spatial gradients define vectors $\boldsymbol{G}_i=(\partial p_1/\partial x_i,\partial p_2/\partial x_i)$, which ACA combines through pseudo-Stokes parameters:
\begin{equation}
 I=\sum_i(G_{1,i}^2+G_{2,i}^2),\quad Q=\sum_i(G_{1,i}^2-G_{2,i}^2),\quad U=\sum_i2G_{1,i}G_{2,i}.
\end{equation}
The principal axis of the second moment of these gradient vectors has orientation $\theta=\frac12\operatorname{atan2}(U,Q)$, giving the local density exponent $\kappa_\rho=\tan\theta=d\log\neff/d\log\leff$ \citep{Zhao2026a}.

\subsection{Multi-threshold KDE}\label{app:kde}
We construct column-density-weighted Gaussian KDEs in the $(\log\neff,\kappa_\rho)$ plane for three nested samples, $\NH>1,2,3\times10^{22}\,\mathrm{cm}^{-2}$, using a bandwidth of 0.2. The density exponent is restricted to $-3<\kappa_\rho<-1$. For each threshold, the KDE peak is defined as the location of the maximum KDE density in the $(\log_{10}\neff,\kappa_\rho)$ plane. We quantify its shift between the lowest and highest column-density thresholds as
\begin{equation}
 \Delta_{\rm KDE}=\left[(\Delta\log_{10}\neff)^2+(\Delta\kappa_\rho)^2\right]^{1/2}.
\end{equation}

\subsection{Core Catalog and Infrared Measurements}\label{app:ir}
Core positions and masses are taken from the Cygnus-X catalog of \citet{LiQiu2026}, extracted uniformly with \texttt{getsf} from the Cygnus-X column-density map. Sources are selected by position inside the same DR21 and W75N regional masks used in the tidal analysis. We use $20\,\Ms$ as the high-mass threshold following that catalog and $50\,\Ms$ to identify the most massive cores.

For the infrared comparison, the 8, 24, and 70~$\mu$m maps are converted to common surface-brightness units of MJy~sr$^{-1}$ and reprojected to the common WCS. For each region and band, the local background $I_{\nu,\rm bg}$ is the sigma-clipped median in a rectangular annulus extending 25 pixels beyond the regional box. For finite pixels inside the region, the background-subtracted integrated flux is
\begin{equation}
 F_\nu=\Omega_{\rm pix}\sum_p(I_{\nu,p}-I_{\nu,\rm bg}),
\end{equation}
with surface-brightness units converted consistently before integration. We define $I_{\nu,99}$ as the 99th percentile of the corresponding background-subtracted intensities and the complementary cumulative area fraction as $A(>I_\nu)/A_{\rm region}$, where $A_{\rm region}$ is the total area of the regional mask. The 8 and 24 $\mu$m maps are the IRAC and MIPS mosaics of the \textit{Spitzer} Cygnus-X Legacy Survey (IRSA dataset DOI: \href{https://doi.org/10.26131/IRSA402}{10.26131/IRSA402}), while the 70 $\mu$m map is the \textit{Herschel}/PACS map from the CENSUS data set \citep{Cao2019}.

\section{Reference Tidal States with Increasing Anisotropy}\label{app:states}
\begin{figure*}[htbp]
 \centering\includegraphics[width=0.9\textwidth]{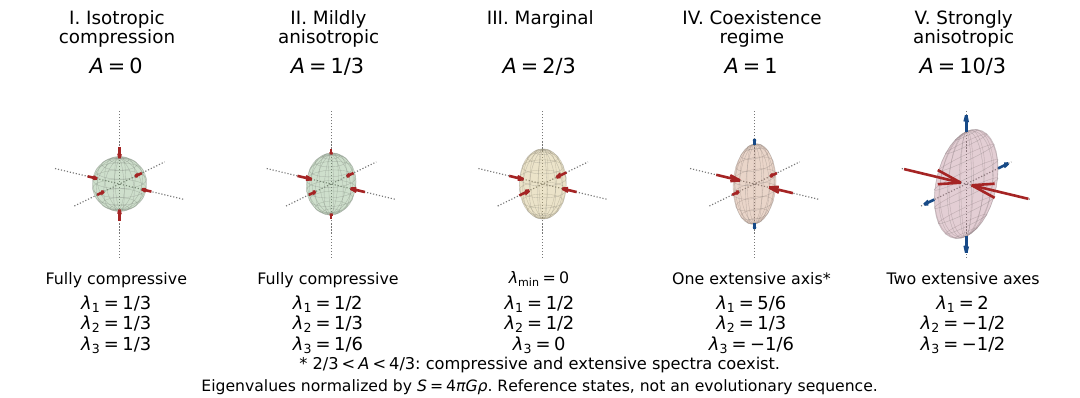}
 \caption{\textbf{Reference tidal states with increasing anisotropy.} Each panel shows a reference eigenvalue spectrum normalized by $S=4\pi G\rho$ ($\lambda_1+\lambda_2+\lambda_3=1$ in normalized units), with arrows along each principal axis (inward for compression, outward for extension, length proportional to $|\lambda_i|$). The labels $\lambda_1,\lambda_2,\lambda_3$ identify the three schematic axes and are not ordered. I. Isotropic compression ($A=0$). II. Mildly anisotropic ($A=1/3$): fully compressive with a preferred axis. III. Marginal ($A=2/3$): one eigenvalue vanishes. IV. Coexistence ($A=1$): compressive and extensive spectra coexist for $2/3<A<4/3$; the panel shows one extensive configuration. V. Strongly anisotropic reference state ($A=10/3$): two directions are extensive; the local threshold implies $\texten<\tff$ at the same position (Appendix~\ref{app:bound}).}\label{fig:states}
\end{figure*}

The tidal anisotropy $A$ (Equation~\ref{eq:A}) traces a continuous sequence of reference tidal states, illustrated schematically in Figure~\ref{fig:states} for eigenvalues normalized by $S=4\pi G\rho$. In the schematic, $\lambda_1,\lambda_2,\lambda_3$ label the three illustrated principal axes and are not ordered; throughout the analysis we use the ordered eigenvalues $\lambda_{\min}\leq\lambda_{\rm mid}\leq\lambda_{\max}$.

At $A=0$, the tidal field is perfectly isotropic: all three eigenvalues equal $S/3$, and the gas experiences uniform compression. As $A$ increases, the eigenvalue spectrum spreads. A vanishing minimum eigenvalue first becomes attainable at $A=2/3$, through the marginal ordered spectrum $(\lambda_{\min},\lambda_{\rm mid},\lambda_{\max})/S=(0,1/2,1/2)$; an extensive direction becomes possible for $A>2/3$. In the range $2/3<A<4/3$, fully compressive and extensive spectra coexist at the same $A$, because the anisotropy alone does not determine the sign of the smallest eigenvalue. At $A=4/3$ a non-extensive spectrum is possible only at the marginal limit $(0,0,1)$; for $A>4/3$ at least one eigenvalue must be negative.

The threshold adopted in this work, $\Ah=10/3$, is represented by the ordered reference spectrum $(\lambda_{\min},\lambda_{\rm mid},\lambda_{\max})/S=(-1/2,-1/2,2)$. In this reference state two of the three principal directions are extensive, and the compressive eigenvalue exceeds the isotropic reference by a factor of six. The number of extensive axes depends on the eigenvalue spectrum: $(-1.2,0.2,2.0)$ also has $A=10/3$ but only one negative eigenvalue. The reference spectra in Figure~\ref{fig:states} illustrate a continuous range of tidal anisotropy. As shown in Appendix~\ref{app:bound}, a local value $A\geq10/3$ implies that the stretching time is shorter than the free-fall time at the same position.

\section{Local Anisotropy and the Tidal-to-Free-Fall Timescale Ratio}\label{app:bound}
At a given position in an extensive region, let $S=4\pi G\rho$ and $q=-\lambda_{\min}/S>0$, with both timescales evaluated using the tidal field and density at that position. Then $\Rt^2=8/(3\pi^2q)$. For ordered eigenvalues with positive trace, $A\leq4/3+4q$, so $A\geq10/3$ requires $q\geq1/2$, giving $\Rt\leq(16/3\pi^2)^{1/2}\simeq0.735<1$. This is a local bound. A large mass-weighted projected anisotropy $\Aproj$ does not require $A$ to exceed the threshold at the density peak. The selection $\Aproj\geq\Ah$ in Section~\ref{sec:timescale} therefore identifies columns with strong projected anisotropy; their timescale ratios at $z=0$ are determined separately.

\section{Density-Exponent Spatial Maps}\label{app:density}
\begin{figure*}[htbp]
 \centering\includegraphics[width=0.96\textwidth]{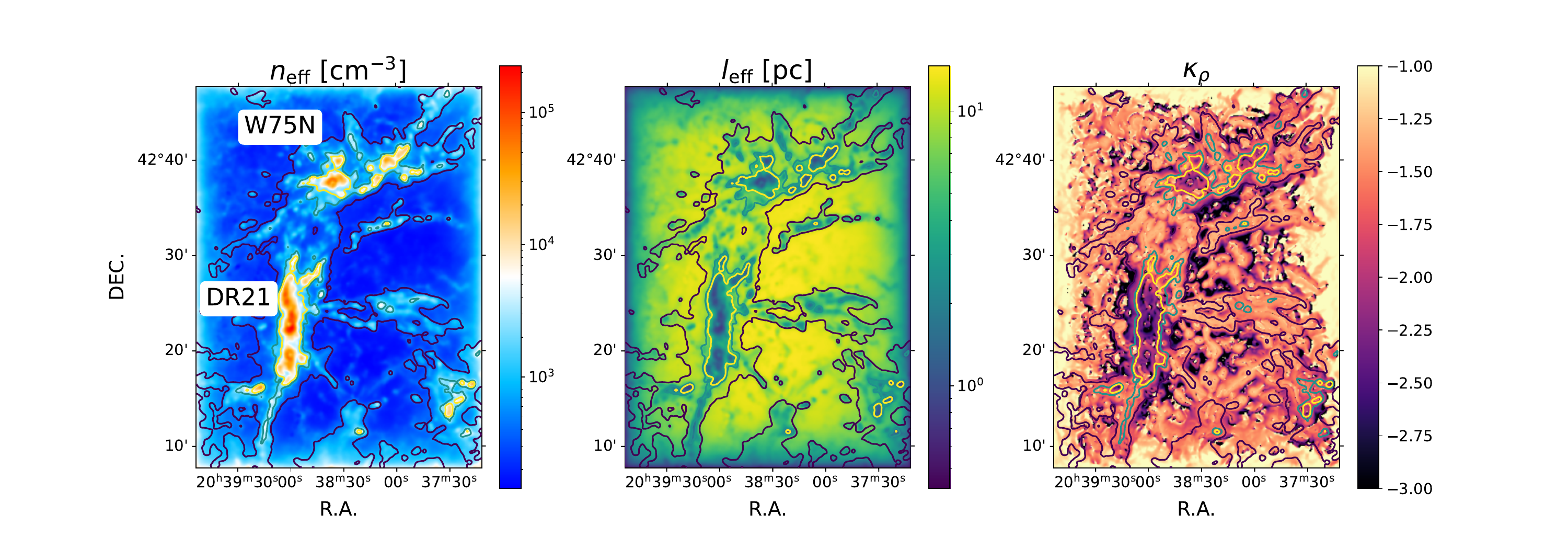}
 \caption{\textbf{VDM-derived density properties.} Left to right: effective molecular-hydrogen number density $\neff$, effective thickness $\leff$, and density exponent $\kappa_\rho$. Contours: $\NH=(1,2,3)\times10^{22}\,\mathrm{cm}^{-2}$. DR21 is dominated by a continuous narrow ridge with steep density exponents; W75N contains several nested structures.}\label{fig:density}
\end{figure*}

The effective molecular-hydrogen number density, effective thickness, and density exponent are shown in Figure~\ref{fig:density}.

\end{document}